\documentclass{aa}

\usepackage{txfonts}
\usepackage{multirow}
\usepackage{hyperref}
\usepackage{graphicx}
\usepackage{fontawesome}
\usepackage{xcolor}

\graphicspath{{figures/}}

\definecolor{myblue}{HTML}{387EE1}
\definecolor{mygreen}{HTML}{00964B}
\definecolor{myred}{HTML}{D9321F}

\hypersetup{
        colorlinks=true,
        linkcolor=myred,
        urlcolor=myblue,
        citecolor=mygreen
}

\newcommand{\quotes}[1]{``#1''}

\begin{document}

\title{Photometric selection and redshifts for quasars\\ in the Kilo-Degree Survey Data Release 4\thanks{We publicly release the catalog at \url{kids.strw.leidenuniv.nl/DR4/quasarcatalog.php} and the code at \url{github.com/snakoneczny/kids-quasars}.}\fnmsep\thanks{A copy of the catalog is available at the CDS via the anonymous ftp to \url{cdsarc.u-strasbg.fr} (\url{130.79.128.5}) or via \url{cdsweb.u-strasbg.fr/cgi-bin/qcat?J/A+A/}.}}

\author{
        S.J.~Nakoneczny\inst{\ref{ncbj}} \and
        M.~Bilicki\inst{\ref{cft}} \and
        A.~Pollo\inst{\ref{ncbj},\ref{uj}} \and
        M.~Asgari\inst{\ref{edinburgh}} \and
    A.~Dvornik\inst{\ref{bochum}} \and
    T.~Erben\inst{\ref{bonn}} \and
    B.~Giblin\inst{\ref{edinburgh}} \and
    C.~Heymans\inst{\ref{edinburgh},\ref{bochum}} \and
    H.~Hildebrandt\inst{\ref{bochum}} \and
    A.~Kannawadi\inst{\ref{princeton}} \and
    K.~Kuijken\inst{\ref{leiden}} \and
    N.R.~Napolitano\inst{\ref{sun_yat-sen}} \and
    E.~Valentijn\inst{\ref{groningen}}
}

\institute{
        National Centre for Nuclear Research, Astrophysics Division, ul. Pasteura 7, 02-093 Warsaw, Poland \label{ncbj} \and
        Center for Theoretical Physics, Polish Academy of Sciences, al. Lotnik\'{o}w 32/46, 02-668 Warsaw, Poland \label{cft} \and
        Astronomical Observatory of the Jagiellonian University, 31-007 Krak\'{o}w, Poland \label{uj} \and
        Institute for Astronomy, University of Edinburgh, Royal Observatory, Blackford Hill, Edinburgh, EH9 3HJ, U.K. \label{edinburgh} \and
        Ruhr University Bochum, Faculty of Physics and Astronomy, Astronomical Institute (AIRUB), German Centre for Cosmological Lensing, 44780 Bochum, Germany \label{bochum} \and
        Argelander-Institut für Astronomie, Auf dem Hügel 71, 53121 Bonn / Germany \label{bonn} \and
        Department of Astrophysical Sciences, Princeton University, 4 Ivy Lane, Princeton, NJ 08544, USA \label{princeton} \and
        Leiden Observatory, Leiden University, P.O.Box 9513, 2300RA Leiden, The Netherlands \label{leiden} \and
        School of Physics and Astronomy, Sun Yat-sen University, Guangzhou 519082, Zhuhai Campus, P.R. China \label{sun_yat-sen} \and
    Kapteyn Institute, University of Groningen, PO Box 800, NL 9700 AV Groningen \label{groningen}
}

\authorrunning{S.J.~Nakoneczny et al.}

\offprints{S.J.~Nakoneczny, \email{\url{szymon.nakoneczny@ncbj.gov.pl}}.}

\abstract{
We present a catalog of quasars with their corresponding redshifts derived from the photometric Kilo-Degree Survey (KiDS) Data Release 4. We achieved it by training machine learning (ML) models, using optical \textit{ugri} and near-infrared $ZYJHK_s$ bands, on objects known from Sloan Digital Sky Survey (SDSS) spectroscopy. We define inference subsets from the 45 million objects of the KiDS photometric data limited to 9-band detections, based on a feature space built from magnitudes and their combinations. We show that projections of the high-dimensional feature space on two dimensions can be successfully used, instead of the standard color-color plots, to investigate the photometric estimations, compare them with spectroscopic data, and efficiently support the process of building a catalog. The model selection and fine-tuning employs two subsets of objects: those randomly selected and the faintest ones, which allowed us to properly fit the bias versus variance trade-off. We tested three ML models: random forest (RF), XGBoost (XGB), and artificial neural network (ANN). We find that XGB is the most robust and straightforward model for classification, while ANN performs the best for combined classification and redshift. The ANN inference results are tested using number counts, Gaia parallaxes, and other quasar catalogs that are external to the training set. Based on these tests, we derived the minimum classification probability for quasar candidates which provides the best purity versus completeness trade-off: $p(\rm QSO_{cand}) > 0.9$ for $r<22$ and $p(\rm QSO_{cand}) > 0.98$ for $22 < r < 23.5$. We find 158,000 quasar candidates in the safe inference subset ($r < 22$) and an additional 185,000 candidates in the reliable extrapolation regime ($22 < r < 23.5$). Test-data purity equals 97\% and completeness is 94\%; the latter drops by 3\% in the extrapolation to data fainter by one magnitude than the training set. The photometric redshifts were derived with ANN and modeled with Gaussian uncertainties. The test-data redshift error (mean and scatter) equals $0.009 \pm 0.12$ in the safe subset and $-0.0004 \pm 0.19$ in the extrapolation, averaged over a redshift range of $0.14 < z < 3.63$ (first and 99th percentiles). Our success of the extrapolation challenges the way that models are optimized and applied at the faint data end. The resulting catalog is ready for cosmology and active galactic nucleus (AGN) studies.
}


\maketitle

\section{Introduction}

Object type and redshift or radial velocity are basic observables in astronomy. They can be precisely determined based on emission and absorption lines from spectroscopy, but they are more difficult to extract from photometric broad-band surveys. However, photometric surveys are often the only feasible approach, particularly for large-scale structure (LSS) studies, which require a high number density and completeness as well as samples of millions of objects. Upcoming large photometric surveys, such as the Vera Rubin Observatory Legacy Survey of Space and Time \citep[LSST,][]{Ivezic:2019}, will provide an unprecedented number of objects and depth of observations.

Quasars (QSOs) stand out as some of the most distant objects we can observe. Unlike regular galaxies, these extragalactic sources cannot be easily identified based on their angular sizes because similarly to stars, they are mostly point-like. We observe QSOs up to very high redshifts because of the accretion of matter on supermassive black holes \citep{Kormendy:2013}, which leads to  enormous amounts of energy being radiated out.  Quasars are important for LSS studies as they reside in dark matter halos of masses above $10^{12} M_{\odot}$ \citep{Eftekharzadeh:2015, DiPompeo:2016}, which makes them highly biased tracers of the LSS \citep{DiPompeo:2014, Laurent:2017}. Possible applications of QSOs in cosmology include tomographic angular clustering \citep{Leistedt:2014, Ho:2015}, the analysis of cosmic magnification \citep{Scranton:2005}, measurement of halo masses \citep{DiPompeo:2017}, cross-correlations with various cosmological backgrounds \citep{Sherwin:2012, Cuoco:2017, Stolzner:2018}, and even the calibration of the reference frames for Galactic studies \citep{Lindegren:2018}.

At any cosmic epoch, QSOs are sparsely distributed in comparison to inactive galaxies. Therefore, wide-angle surveys are essential to obtain catalogs containing a sufficient number of  QSOs to be useful for studies where good statistics are important. Previous spectroscopic surveys, such as the 2dF QSO Redshift Survey \citep[2QZ,][]{Croom:2004} or the Sloan Digital Sky Survey \citep[SDSS,][]{York:2000, Lyke:2020}, provided  $\sim 10^4$-$10^5$ QSOs. In spectroscopy, QSO detection and redshift measurement are based on broad emission lines such as [OIII]$\lambda 5007$/H$\beta$, [NII]$\lambda$6584/H$\alpha$ \citep{Kauffmann:2003, Kewley:2013}. Many surveys exploit this approach, including: 2QZ, 2dF-SDSS LRG, and QSO \citep[2SLAQ,][]{Croom:2009}, SDSS, or the forthcoming DESI \citep{DESI:2016} and 4MOST \citep{deJong:2019, Merloni:2019, Richard:2019}.

Spectral energy distribution (SED) fitting is a standard approach to analyze photometry of galaxies with active galactic nuclei (AGN), which include QSOs in particular. It allows one to derive physical properties \citep{Ciesla:2015, Stalevski:2016, Calistro:2016, Yang:2020, Malek:2020} and estimate photo-zs \citep{Salvato:2009, Salvato:2011, Fotopoulou:2016, Fotopoulou:2018}. The QSO selection in photometry is commonly based on color-color cuts \citep{Warren:2000, Maddox:2008, Edelson:2012, Stern:2012, Wu:2012, Secrest:2015, Assef:2018}. More sophisticated and arguably more robust approaches to QSO selection are the probabilistic methods \citep{Richards:2004, Richards:2009a, Richards:2009b, Bovy:2011, Bovy:2012, DiPompeo:2015, Richards:2015}, while machine learning (ML) has been gaining popularity in this respect as well \citep{Brescia:2015, Carrasco:2015, Kurcz:2016, Nakoneczny:2019, Logan:2020}. Machine learning models have also been applied to derive QSO photometric redshifts \citep[photo-zs,][]{Brescia:2013, Yang:2017, Pasquet:2018, Curran:2020}.

In the context of the Kilo-Degree Survey \citep[KiDS,][]{deJong:2013}, which is the focus of our paper, the QSO-related studies have so far dealt with high-redshift ($z\sim6$) QSOs \citep{Venemans:2015}, heavily reddened QSOs \citep{Heintz:2018}, and selecting QSOs to search for strong-lensing systems \citep{Spiniello:2018, Khramtsov:2019}, while in \citet[][hereafter N19]{Nakoneczny:2019} we present an ML QSO detection analysis in KiDS Data Release 3 (DR3, \citealt{deJong:2017}). We note that, in general, every QSO present in KiDS multiband catalogs has a redshift estimate derived with the Bayesian Photometric Redshift code (\textsc{bpz}, \citealt{Benitez:2000}), as such photo-zs are computed by default for each cataloged object. However, these redshifts are usually not correct for QSOs as their derivation is optimized at galaxies used for weak lensing studies \citep{Kuijken:2015} and in particular proper AGN templates are not used in the \textsc{bpz} implementation. Similarly, the KiDS database does not offer any direct indication of which sources could potentially be QSOs.

In our previous work (N19), we performed a classification in KiDS DR3, using optical $ugri$ broad-band data. The random forest (RF) achieved QSO purity of 91\% and completeness of 87\%. The failures in QSO classification -- mislabeling them as stars -- occurred mostly at a QSO redshift of $2 < z < 3$. Due to the magnitude limit of training data available from SDSS, we restricted the catalog to $r < 22$. This resulted in 190,000 QSO candidates, which were selected from 3.4 million objects taken as the inference data from KiDS DR3 based on four broad-band detections and data quality considerations.

In this paper, we perform classification and redshift estimation using optical and near-infrared (near-IR)\ broad-bands of KiDS DR4 \citep{Kuijken:2019}, which incorporates the partner VISTA Kilo-degree Infrared Galaxy \citep[VIKING,][]{Edge:2013} measurements. Our main goal is to create a catalog of QSOs, optimized for the highest purity and completeness, with robust photometric redshift estimates. We test what near-IR imaging brings to classification in terms of separating QSOs from stars. We aim to fit ML models for the best bias versus variance trade-off in order to achieve reliable results at the faint data end, not represented well by the spectroscopic data used in training. We verify whether randomly selected subsets of spectroscopic objects used to test ML models lead to the proper bias-variance trade-off, or if it is better to also validate based on the faintest objects, which are never seen during training. This is necessary to assess the level of overfitting, address the problem of extrapolation in the feature space (a space of n-dimensional feature vectors consisting of, for instance, magnitudes and colors), and provide reliable estimates at the faint data end. We test different strategies of building features from broad-band magnitudes, find which of the most popular ML models perform best for classification and redshifts, and model QSO photometric redshift uncertainties with a Gaussian output layer in an Artificial Neural Network (ANN). Last but not least, we check whether projection of high-dimensional space on two dimensions (2D) can substitute the standard color-color plots as a tool to inspect the feature space coverage and differences between spectroscopic and photometric results, and to meaningfully interpret the data.

The paper is organized as follows. In Section \ref{section_methodology} we describe the data and the methodology for QSO selection, redshift estimation, extrapolation in the feature space, and bias-variance tuning; in Section \ref{section_results} we provide results of experiments done on a cross-match with spectroscopic data, properties of the final catalog, and purity-completeness calibration; in Section \ref{section_discussion} we discuss the main findings, strengths, and weaknesses of the approach, and we outline possible extensions. Where relevant, we use the flat $\Lambda$CDM cosmology based on the Nine-Year Wilkinson Microwave Anisotropy Probe \citep[WMAP9,][]{Hinshaw:2013} with $H_0 = 69.3$ km/s/Mpc and $\Omega_m = 0.287$.

\section{Data and methodology}
\label{section_methodology}

\subsection{Data}
\label{subsection_data}

KiDS\footnote{\url{http://kids.strw.leidenuniv.nl}} is an optical wide-field imaging survey with the OmegaCAM camera \citep{Kuijken:2011} at the VLT Survey Telescope \citep[VST,][]{Capaccioli:2012}, specifically designed for measuring weak gravitational lensing by galaxies and a large-scale structure \citep{Joudaki:2017, vanUitert:2018, Asgari:2020, Heymans:2020, Hildebrandt:2020-shear, Wright:2020}. It consists of 1350 square degrees imaged in four broad-band \textit{ugri} filters. The current fourth data release \citep{Kuijken:2019} is the penultimate one; it covers a total of 1006 deg$^2$ and provides a list of $\sim100$ million (100M) objects based on the \textit{r}-band detections. It also includes $ZYJHK_s$ photometry from the partner VIKING. The mean limiting AB magnitude (5 $\sigma$ in a 2 arcsec. aperture) of KiDS is $\sim25$ in the \textit{r} band. The optical depth, wide sky coverage, and multiwavelength imaging make this survey an ideal resource for QSO science.

Additional discriminatory power for QSO selection and photo-zs could be provided from mid-infrared bands, such as from the Wide-field infrared Survey Explorer \citep[WISE, ][]{Wright:2010}, as shown for instance in \cite{Logan:2020}. However, in this work we decide to limit the selection to the 9-band KiDS+VIKING only, as adding the WISE data would severely limit our dataset. For instance, at the $r < 23.5$ KiDS limit, only $\sim$19\% of the sources have a counterpart in WISE within a 3" matching radius. This fraction decreases even further with an increased KiDS depth.

We solved the problem of QSO detection with classification models and derived photo-zs with regression models. Reliably applied supervised machine learning requires data from which it can learn a solution to the problem and assurance that the inference data are well represented by its training subset. We created one feature set for both classification and regression to keep the models consistent in predictions. We limited the KiDS data to 9-band detections (sources which have all the nine bands measured) in order to provide the most reliable set of features (Section \ref{subsection_features}). During the experiments, we calculated all of the scores, including completeness, in the limited set of 9-band detections, but the number counts of the final catalog compare completeness with respect to all possible QSOs. The feature set includes nine magnitudes derived with the Gaussian Aperture and PSF (GAaP) photometry method \citep{Kuijken:2008}, 36 colors, 36 ratios of every magnitude pair, and the following two morphological classifiers: SExtractor-based CLASS\_STAR \citep[called the stellarity index here;][]{Bertin:1996}, and the third bit of SG2DPHOT -- KiDS star versus galaxy separation flag based on source $r$-band morphology\footnote{Flag values are: 1 (high-confidence star candidates), 2 (objects with FWHM smaller than stars in the stellar locus), 4 (stars according to S/G separation), and 0 otherwise (galaxies); flag values are summed. See sect.\ 4.5.1 of \cite{deJong:2015} for details.} \citep{deJong:2015, deJong:2017}. In Section \ref{subsection_features} we describe the experiments which led to this final set of 83 features. The 9-band detection requirement reduces the number of objects from $\sim$100M to $\sim$~45M, which creates the inference set.

The training set was derived from cross-matching the inference data with the Sloan Digital Sky Survey DR14 \citep[SDSS,][]{Abolfathi:2018} spectroscopic observations\footnote{More recent SDSS DR16 does not provide additional overlap with KiDS with respect to DR14.}. The SDSS survey provides three basic classes: galaxies, QSOs, and stars, which we use to define a three-class classification problem. After removing objects flagged with warnings by SDSS, we obtained a training subset of 152k objects (69\% galaxies, 11\% QSOs, 20\% stars). The training set is limited to $r\sim22$ by SDSS (99\% of training is at $r<21.98$), which is about three magnitudes brighter than the depth of the KiDS inference data. The results of machine learning predictions for $r\gtrsim22$ may be incorrect due to the resulting extrapolation in the feature space.

\subsection{Inference subsets}
\label{subsection_inference_subsets}

\begin{figure}
        \resizebox{\hsize}{!}{\includegraphics{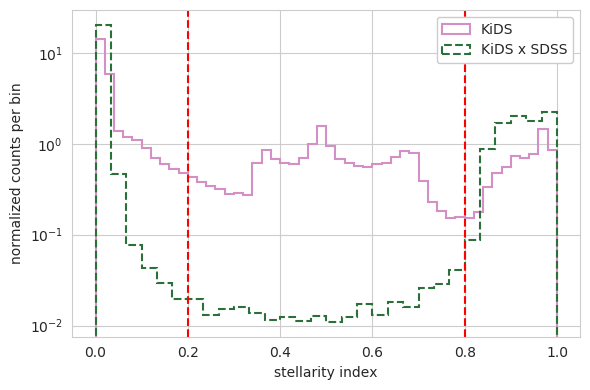}}
        \caption{Normalized histograms of the CLASS\_STAR stellarity index in the training (KiDS x SDSS) and inference (KiDS) datasets. The intermediate values represent failures of the morphological classifier. Those values are not commonly present in the training data, thus we cannot expect the ML models to work correctly for objects with such index values. We consider the sources in between the red dashed lines as unsafe for the inference.}
        \label{fig:class_star}
\end{figure}

In this section we define the inference subsets based on feature set considerations. The training set we use is a small subset of the KiDS inference data and does not fully cover the feature space. An inference on parts of the feature space not covered by the training data may result in the deterioration of results or a complete failure, due to new  combinations of features or completely new feature values. For continuous features, such as magnitude, we may expect well-generalized models to extrapolate with deteriorating quality of the estimations. In case of discrete features, whose new values cannot be understood based on the ones available in training, supervised ML models may fail completely. We therefore define inference subsets based on how feature coverage changes from training to inference data and how this can affect the ML models.

The morphological classifiers tend to fail at the faint data end. We used them to achieve the highest accuracy at the bright end, and as a proxy for data quality at the faint end. The SExtractor-based stellarity index has a continuous distribution between zero and one, with large values indicating point-like objects, small values corresponding to extended objects, and intermediate values pointing to classifier failure. Because the failures are almost not present in the bright training data, ML models do not  understand their meaning (Fig. \ref{fig:class_star}). We therefore only consider the stellarity index ranges $(0, 0.2)$ and $(0.8, 1)$ covered by the training data as safe for the inference. Choosing cuts which admit more objects to the safe inference subset might increase completeness at the cost of purity, while stricter cuts do the opposite. The second morphological classifier we used, SG2DPHOT, is a discrete one, whose first and third bits indicate stars, and its failure is indicated by the zero value, which is the same as for galaxies. We find empirically that using only its third bit provides the best improvement in our results. Cleaning the uncertain stellarity index values removes most of the SG2DPHOT failures.

The magnitude range $r < 22$ is covered by the training data, whereas for $r > 22$ we expect ML models to extrapolate with deteriorating quality. We define three inference subsets based on the feature space coverage, morphological classification quality, and the $r$-band depth of the survey. Firstly, the safe subset is $r < 22$ and a stellarity index of $\notin  (0.2, 0.8)$; secondly, the extrapolation subset is $r \in (22, 25)$ and a stellarity index of $\notin (0.2, 0.8)$; and lastly, the unsafe subset is $r > 25$ or a stellarity index of $\in (0.2, 0.8)$.

\begin{figure*}
    \sidecaption
        \includegraphics[width=12cm]{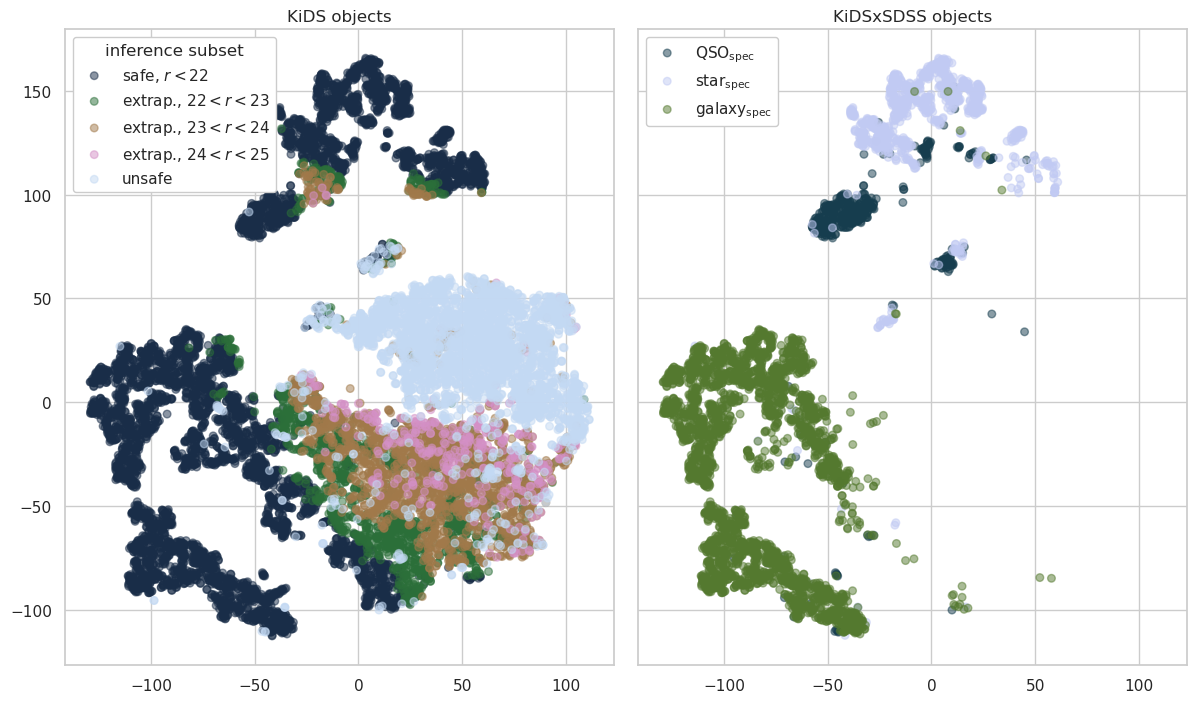}
        \caption{t-SNE projections. \textit{Left:} Inference subsets. \textit{Right:} SDSS spectroscopic classification. The visualizations were made on subsets of 12k objects. The real density of objects at any part of the feature space is 3.8k times higher than visualized. We can see three main groups. The point-like objects cover the top part, extended ones are located at the bottom, and those with undetermined morphology are placed in the middle, in the unsafe subset. The spectroscopic data cover only the bright part of the photometric data; this illustrates the extrapolation problem to address with machine learning. The results of the inference are later investigated on similar plots (Section \ref{subsection_final_catalog}), which we consider a more robust approach than investigating color-color diagrams.}
        \label{fig:t-SNE_subsets}
\end{figure*}

We visualize the KiDS feature space and the inference subsets with t-distributed Stochastic Neighbor Embedding \citep[t-SNE,][]{Maaten:2008} in Fig.~\ref{fig:t-SNE_subsets}. t-SNE belongs to a family of manifold learning algorithms, and it allows us to visualize high dimensional and nonlinear data structures with much simpler two dimensional embeddings. We created the visualization with the same set of 83 features that are used in classification and redshift estimation in order to visualize the same feature space. Due to the computational complexity of t-SNE, we took 8k random objects from KiDS data and merged them with 4k random objects from KiDSxSDSS cross-match to visualize the spectroscopic classes, which are sparse in the whole KiDS data, and put emphasis on the much fainter inference data. The plots show the main groups of spectroscopic classes and their placement over the whole feature space. The safe subset at $r<22$ matches the part of the feature space covered by the training data, confirming that a single cut on the \textit{r} magnitude assigns proper limits to the other magnitudes, colors, and ratios; we observed the same result previously in N19, where we matched only the $ugri$ magnitudes, colors, and ratios. The main star and QSO groups are separated in the training data, but not in the whole KiDS data. The first part of the extrapolation subset at $r \in (22, 23)$ is located close to the training data, and it may provide reliable estimations. The rest of the extrapolation set covers fainter and more complicated parts of the feature space, such as the joining space between QSO and star groups at $23 < r < 24$, thus such objects have a lower chance of their classification predictions being correct.

We used the 2D visualization to investigate estimation performance of the ML models. The models work with highly dimensional data, which makes it difficult to visualize the decision boundaries. We did not investigate the color-color plots due to the large number of possible combinations and the required domain knowledge of how to interpret them. Instead, the manifold learning, such as t-SNE, visualizes nonlinear data structures and this allows us to understand the models as well as, or better than, it would be possible with the color-color plots. Additionally, we used the embedding to have insight into the extrapolation part of the feature space, which cannot be tested with methods based on ground-truth data.

\begin{figure}
        \resizebox{\hsize}{!}{\includegraphics{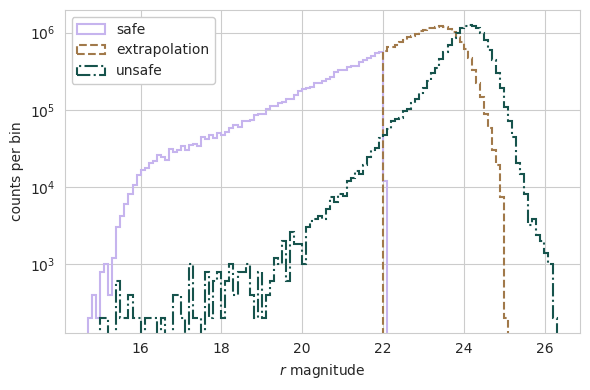}}
        \caption{Distribution of inference subsets over the \textit{r} magnitude. The limit of the SDSS training data, $r = 22$, defines the lower limit of the extrapolation subset. The morphological classifier failure and sources beyond survey depth ($r>25$) provide the unsafe subset (fig. \ref{fig:class_star}). The extrapolation subset is complete up to $r < 23.5$. The safe subset covers 21\% of data, extrapolation 45\%, and unsafe 34\%.}
        \label{fig:inference_subsets}
\end{figure}

Figure \ref{fig:inference_subsets} shows $r$ magnitude distributions for the inference subsets. The safe subset was cut at $r=22$, while the extrapolation and unsafe subsets overlap in magnitudes. We can see that the extrapolation subset is complete to $r<23.5$, which puts a completeness limit on our catalog. We expect that the number counts of QSOs identified using the currently available training sets would become incomplete at $r>23.5$.

\subsection{Validation procedure}
\label{subsection_evaluation}

\begin{figure*}
        \centering
        \resizebox{\hsize}{!}{\includegraphics{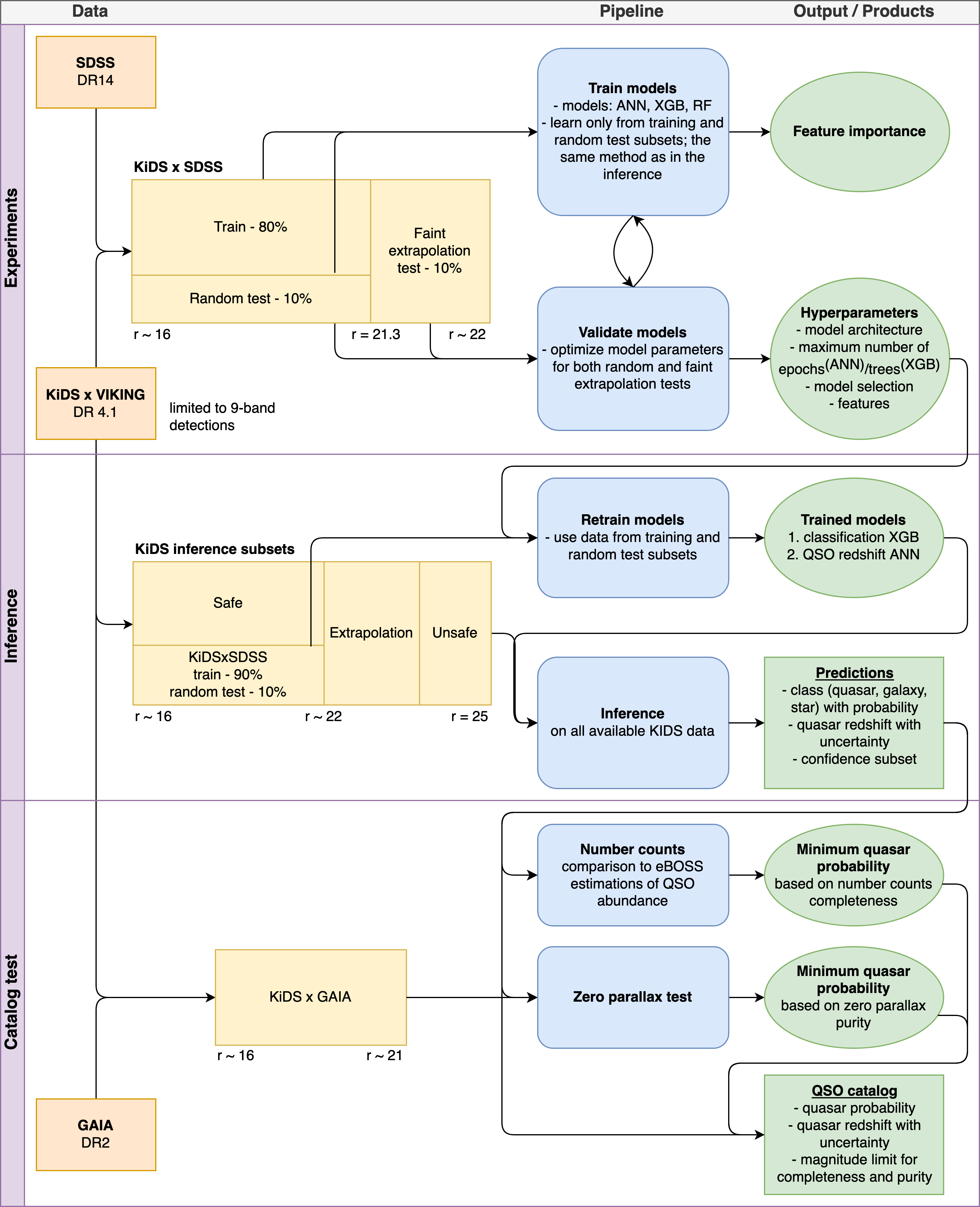}}
        \caption{Methodology diagram. The procedure consists of three main parts: experiments as well as inference and catalog tests. The experiments are based on the cross-match between KiDS and SDSS data, and they include the repeatable process of training and evaluating ML models. The training is based only on the train and random test subsets, while the hyper-parameter tuning uses both random and faint extrapolation tests. The best hyper-parameters found are used in the inference to train new models, now on the whole range of magnitudes available in the training data. The raw predictions were then tested with number counts and Gaia parallaxes to calibrate the final catalog with probability cuts for the optimal purity-completeness trade-off.}
        \label{fig:methodology}
\end{figure*}

Proper design of the testing methods is one of the main goals of this paper so as to make sure we did not overfit the models. Validation data have to differ from the training set to ensure proper model generalization. A randomly chosen sample of data which densely covers the feature space might not fully show the overfitting effects, and this might have a very negative influence on the inference at the faint data end, both for classification and photo-zs. We used additional spectroscopic surveys to introduce some differences from the training data and tested the final predictions (Section \ref{subsection_external_test}). During the experiments, we used internal data characteristics to differentiate training from validation. The approach is similar to time series processing, where validation data should consist of dates later than the training ones. Similarly, we chose the faintest objects to test the regularization of the models. Another option would be to use highest-redshift objects, chosen separately for each class as they reside at different ranges of redshift, which would test the prediction of values not seen during training. However, radial velocities of stars measured by SDSS obviously do not correlate with photometry and we would not observe any variation in star colors between the training and validation data. As magnitude correlates with redshift in the case of QSOs and galaxies, we expect the faint test to evaluate the extrapolation accuracy of ML models with respect to the estimated redshift values. Figure \ref{fig:methodology} explains the whole methodology in blocks illustrating experiments as well as inference and catalog testing.

\begin{table*}
        \caption{Train and test subsets of the KiDSxSDSS data. We selected the faintest 10\% of the data as the faint extrapolation test. This splits the training data at $r=21.3$. We used the same amount of objects at $r<21.3$ as in the faint-end for the random test.}
        \centering
        \begin{tabular}{l l c c c c}
                \hline\hline
                & & size & quasar & galaxy & star \\
                \hline
                train & r < 21.3 & 105k & 11k (11\%) & 71k (68\%) & 23k (22\%) \\
                test random & r < 21.3 & 13k & 1.5k (12\%) & 8.8k (67\%) & 2.8k (21\%) \\
                test faint & 21.3 < r < 22 & 13k & 3.3k (25\%) & 7.2k (55\%) & 2.6k (20\%)\\
                \hline
        \end{tabular}
        \label{table:train_test_split}
\end{table*}

Table \ref{table:train_test_split} summarizes the training and validation sets. We selected the faintest 10\% of the training data as a faint extrapolation test, and the same amount of random objects from the rest of the training data as a random test. Both tests allowed us to correctly tune the models for a bias-variance trade-off and check how the estimations deteriorate when we extrapolate to fainter magnitudes. The faint extrapolation test has a higher contribution of QSOs, which adds to differences between the training and validation. The faint extrapolation test sample in the spectroscopic data, at $21.3<r<22$, should not be confused with the faint extrapolation inference data at $r>22$.

We tested QSO redshifts on two subsets: the true spectroscopic QSOs from SDSS and QSO candidates from the output of an ML model. The QSO candidates may contain true stars and galaxies due to misclassification. As we are solving two distinct tasks, classification to identify QSOs and regression to estimate their redshifts, a test of QSO candidates evaluates the consistency between classification and redshift models and it requires both class and redshift to be assigned correctly. This test informs us about the robustness of the final catalog, and we consider redshift errors obtained in the set of QSO candidates as the most important metric for model selection.

We used the following classification metrics\footnote{\url{https://scikit-learn.org/stable/modules/model_evaluation.html}} \citep[scikit-learn,][]{Pedregosa:2011}: accuracy for the three-class classification problem (QSO, galaxy, and star) as well as purity and completeness for QSO detection. For redshifts, we used:
\begin{itemize}
    \item the mean squared error
    \begin{equation}
    \label{eqn:mse}
        MSE = \frac{1}{N}\Sigma(z_{\rm spec,i} - z_{\rm photo,i})^2,
    \end{equation}
    \item R-squared
    \begin{equation}
    \label{eqn:r2}
        R^2 = 1 - \frac{SS_{\rm RES}}{SS_{\rm TOT}} = 1 - \frac{(z_{\rm spec,i} - z_{\rm photo,i}) ^ 2}{(z_{\rm spec,i} - \bar{z}_{\rm spec}) ^ 2}, and
    \end{equation}
    \item the redshift error
    \begin{equation}
    \label{eqn:z_error}
        \delta z = \frac{z_{\rm photo} - z_{\rm spec}}{1 + z_{\rm spec}},
    \end{equation}
\end{itemize}
where $z_{\rm spec}$ is the true spectroscopic redshift, $z_{\rm photo}$ is the predicted photometric redshift, and $\bar{z}_{\rm spec}$ is the mean spectroscopic redshift of a given validation sample.

We performed 100 bootstrap samplings on random and faint extrapolation tests to make sure that the mean standard errors ($\sigma / \sqrt{100}$) are at about 3-4 decimal places depending on the metric. This gives statistical relevance to the precision with which we report the results.

Due to differences between the training and inference data, we used several methods to test the final catalog: number counts, spatial densities, GAIA parallaxes, and comparison with external quasar catalogs. This way we ensure that any decision on model parameters or feature engineering does not lead to issues in the final inference. Those testing methods allowed us to calibrate the purity versus completeness ratio of the final quasar catalog by setting the minimum classification probability. With calibrated classification, the photometric redshifts might be a good approximation of the real redshift distribution, but further calibrations are possible.

\subsection{Model selection}
\label{subsection_model_selection}

We tested three of the most popular ML models: random forest \citep[RF,][]{Breiman:2001}, XGBoost \citep[XGB,][]{Chen:2016}, and artificial neural networks \citep[ANN,][]{Haykin:1998}. We used Python libraries: \textsc{scikit-learn}, \textsc{Tensorflow} \citep{Abadi:2015}, and \textsc{Keras} \citep{Chollet:2015}. The RF and XGB are ensemble models, in which classification or regression is performed using many decision trees. The RF randomizes the trees by choosing a subset of training data and/or features for each tree. The XGB introduces the boosting procedure which favors selection of data points for which the model has the highest errors. Additionally, it uses gradients to approximate and minimize an error function. The ANNs consists of stacked layers of neurons, with nonlinear activation function in each neuron.

We tested two redshift estimation strategies: one model for all the classes and two specialized models trained separately for quasars and galaxies. In case of the specialized models, we assigned zero redshift to stars. We also tested a neural network model with multiple outputs for classification and redshifts, which allowed us to solve both problems with only one model.

\subsection{Feature engineering}
\label{subsection_feature_engineering}

Feature engineering is one of the ways to tune model complexity, and it is widely used in an ML practice \cite[see][chap.~6]{Bishop:2006}. Already in simpler models, such as linear regression, it allows for increased nonlinearity by applying a kernel. In more complicated models, it leads to better adaptation to the given training data and allows them to extract the true patterns from the inference data. As was explicitly shown in our previous work (N19), feature engineering may bring significant improvements to the results of ML analysis. This should not be considered as a limitation of ML models, but it is a consequence of adjusting the bias versus variance trade-off. However, excessive feature engineering can lead to overfitting; therefore, a reliable testing method is required for this approach to work properly and to match its strategy with proper model regularization. We have indeed designed such tests as described in Sec. \ref{subsection_evaluation}.

We kept the feature engineering fairly simple by considering only the input features and, for magnitudes, their simplest combinations: differences (colors) and ratios. The ratios are widely used in ML and also in astronomy \citep{D'Isanto:2018}, and we used them with success in N19. More complex feature engineering is possible, but we found this strategy sufficient to obtain good results, without a risk of overfitting according to our testing methods. We rank features by their feature importance from XGB models, which was calculated as a sum of gain that a given feature provides to a model in all the splits which are made based on that feature.

\section{Results}
\label{section_results}

\subsection{Feature selection}
\label{subsection_features}

The final set of features consists of 83 values: optical $ugri$ and near-IR $ZYJHK_s$ magnitudes, differences (colors) and ratios of every pair of magnitudes, and two morphological classifiers: the stellarity index from SExtractor and the third bit of SG2DPHOT from KiDS. We tested other bits of the SG2DPHOT without an observable improvement in the results. Ellipticity and other apertures were tested in the previous work (N19) and no significant increase in performance was seen. 

\begin{figure*}
        \resizebox{\hsize}{!}{\includegraphics{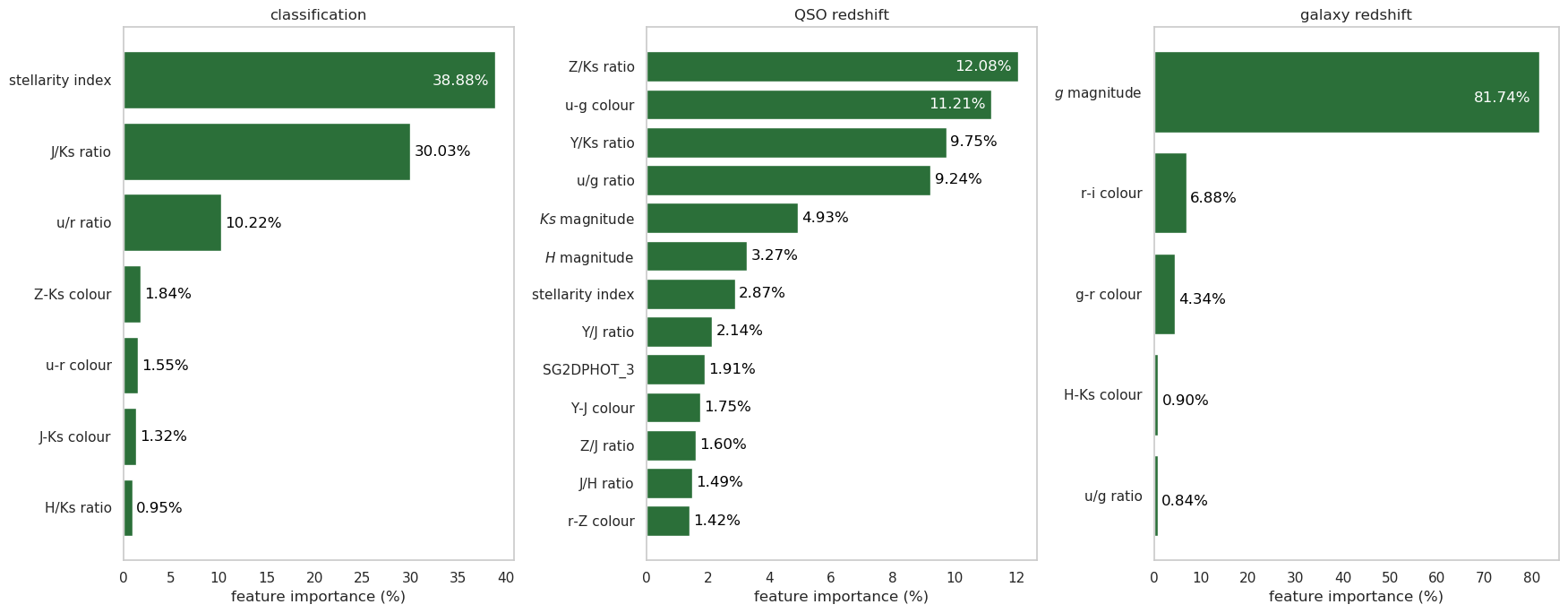}}
        \caption{Feature rankings from the XGB models. \textit{Left:} Classification. \textit{Center:} QSO redshift. \textit{Right:} galaxy redshift. We used the total gain across all splits in which the feature is used. The classification is mostly based on the stellarity index, near-IR $JK_s$, and optical $ur$ bands. The QSO redshifts use all the NIR bands and most of the optical ones, but also the morphological parameters. The galaxy redshifts are based practically only on the optical \textit{gri} magnitudes. Colors and ratios of the same magnitude pairs have a different importance.}
        \label{fig:features_xgb}
\end{figure*}

Figure \ref{fig:features_xgb} shows the most important features for the classification and redshift estimation. We observe the importance of near-IR imaging, which is less affected by dust than the optical bands. The classification is mostly based on colors and magnitude ratios, but the redshift models also use the magnitude values, which is expected due to correlation between apparent magnitude and redshift. Quasar redshifts require more features than galaxy photo-zs, which confirms that they are more challenging to estimate. The most important magnitudes for QSO redshifts, the near-IR $ZK_s$, are the two extreme bands in this range. We observe only one feature of relatively low importance, which mixes the optical and near-IR, the $r-Z$ color. The morphological parameters were also used for QSO redshift, allowing models to distinguish extended low-redshift AGNs.

\begin{table*}
        \caption{Comparison of two feature sets: all 83 features and a limited set of 47 features excluding magnitude ratios, reported on the random ($r < 21.3$) and faint extrapolation ($r \in (21.3, 22)$) tests. The redshifts were tested in two subsets of QSOs: true spectroscopic ones and photometric candidates. The candidates include misclassified sources, e.g., a true star assigned a QSO class and redshift. We mark with bold font the best results, independently for random and faint extrapolation tests. The asterisk (*) marks the approach used to create the final catalog. Recall is the same as completeness, and MSE, R\textsuperscript{2}, and $\delta z$ are given by equations \ref{eqn:mse}, \ref{eqn:r2}, and \ref{eqn:z_error}, respectively.}
        \centering
        \begin{tabular}{c c c c c c c c c c c c}
                \hline\hline
                & & & \multicolumn{3}{c}{classification} & \multicolumn{3}{c}{redshift for true QSOs} & \multicolumn{3}{c}{redshift for QSO candidates} \\
                test & model & features & accuracy & purity & recall & MSE & R\textsuperscript{2} & $\delta z$ & MSE & R\textsuperscript{2} & $\delta z$ \\
                \hline
                random & RF & all features & 99.01\% & 97.39\% & 94.43\% & 0.12 & 85\% & 0.02 $\pm$ 0.14 & 0.12 & 84\% & 0.03 $\pm$ \textbf{0.21} \\
                & & no ratios & 98.98\% & 97.20\% & 94.57\% & 0.11 & 86\% & 0.02 $\pm$ 0.14 & 0.14 & 82\% & 0.04 $\pm$ 0.25 \\
                & XGB & all features & \textbf{99.09\%} & \textbf{97.80\%} & \textbf{95.82\%} & 0.13 & 84\% & 0.02 $\pm$ 0.15 & 0.13 & 83\% & 0.03 $\pm$ \textbf{0.21} \\
                & & no ratios & 99.07\% & \textbf{97.80\%} & 94.66\% & 0.12 & 84\% & \textbf{0.01} $\pm$ 0.15 & 0.12 & 83\% & 0.03 $\pm$ 0.20 \\
                & ANN* & all features* & 98.98\% & 96.90\% & 94.72\% & \textbf{0.09} & \textbf{88\%} & \textbf{0.01 $\pm$ 0.12} & \textbf{0.11} & 85\% & \textbf{0.02} $\pm$ 0.23 \\
                & & no ratios & 98.99\% & 97.58\% & 94.30\% & \textbf{0.09} & \textbf{88\%} & \textbf{0.01 $\pm$ 0.12} & \textbf{0.11} & \textbf{86\%} & \textbf{0.02 $\pm$ 0.21} \\
                \hline
                faint & RF & all features & 97.41\% & 96.07\% & \textbf{92.30\%} & 0.31 & 31\% & 0.02 $\pm$ 0.25 & 0.33 & 31\% & 0.05 $\pm$ 0.38 \\
                extrap. & & no ratios & 97.39\% & 96.12\% & 92.22\% & 0.29 & 35\% & 0.01 $\pm$ 0.24 & 0.31 & 35\% & 0.04 $\pm$ 0.38 \\
                & XGB & all features & 97.41\% & 96.44\% & 92.07\% & 0.27 & 39\% & 0.04 $\pm$ 0.23 & 0.34 & 29\% & 0.08 $\pm$ 0.41 \\
                & & no ratios & \textbf{97.45\%} & 96.55\% & 91.94\% & 0.27 & 39\% & 0.04 $\pm$ 0.22 & 0.33 & 29\% & 0.08 $\pm$ 0.40 \\
                & ANN* & all features* & 97.24\% & 96.54\% & 90.82\% & \textbf{0.22} & \textbf{51\%} & \textbf{0.00 $\pm$ 0.19} & \textbf{0.28} & \textbf{38\%} & 0.04 $\pm$ \textbf{0.37} \\
                & & no ratios & 97.26\% & \textbf{96.94\%} & 90.88\% & 0.24 & 46\% & \textbf{0.00} $\pm$ 0.20 & 0.30 & 32\% & \textbf{0.03} $\pm$ 0.38 \\
                \hline
                \hline
        \end{tabular}
        \label{table:model_comparison}
\end{table*}

Feature importance suggests using ratios of magnitudes, however, the importance is based only on the training set and might fail in showing the effects of overfitting. Table \ref{table:model_comparison} compares the full set of 83 features, with a limited set of 47 features excluding the magnitude ratios. We fine-tuned the models to the full set of 83 features as suggested by the feature importance, but we note that this approach might underestimate the performance of the no-ratio feature set. We observe that the differences between the two feature sets are significant for a faint extrapolation test of photometric redshifts. The ANN trained on the full set of features achieves the best results overall. Due to underestimated performance of the no-ratio feature sets, two scenarios are still possible: the magnitude ratios provide better results on both tests, or lead to overfitting in which case the random test results might be better, but the faint extrapolation results would be worse. It is very important that, to be able to properly assess the possibility of overfitting while using the ratios, the faint extrapolation test is necessary, as the random test already fails to show differences between the two feature sets. In this work, we decided to use the full set of 83 features, suggested by the feature importance. The approach we chose may not be optimal, as more experiments with feature and model engineering are possible. However, as the results we achieved are already very good, more extensive experiments are beyond the scope of this paper.

Additionally, we experimented with reducing the feature set by removing, not whole groups of features (magnitudes, colors, or ratios), but single and least important features used for classification to minimize possible overfitting and increase model interpretability. This provided stable results for classification, but worsened the redshift estimates in the subset of QSO candidates due to lower consistency between the classification and redshift models. The inconsistency between the models results in more objects with either one of the classes or a redshift assigned incorrectly, while the redshifts of the QSO candidates require both the class and the redshift to be assigned correctly.

\subsection{Experiment results}
\label{subsection_sdss_test}

\begin{figure*}
        \resizebox{\hsize}{!}{\includegraphics{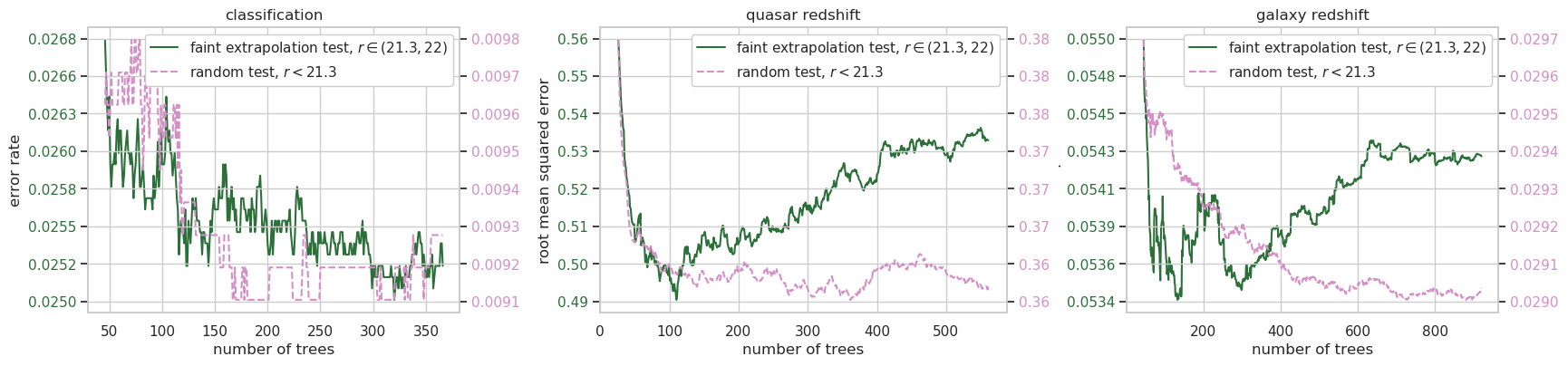}}
        \caption{Learning histories for the XGB models. \textit{Left:} Classification. \textit{Center:} QSO redshift. \textit{Right:} Galaxy redshift. The x-axis shows the number of trees created iteratively during the model training, and the y-axis shows the classification error rate and redshift root mean square error on two different scales for the random and faint extrapolation tests. The errors in the faint test are higher than in the random tests due to extrapolation and higher noise. The models were stopped if the results on the faint test did not improve for 200 consecutive trees. For classification, which is easier to solve than redshift regression, the random test shows minimums sooner, followed by oscillations, while the faint test suggests longer training. For redshifts, which is a more complicated problem, the faint test achieves minimum quickly and then shows overfitting, while the random test suggests longer training.}
        \label{fig:history_xgb}
\end{figure*}

Figure \ref{fig:history_xgb} compares XGB training histories (number of trees used) for the classification and redshifts. The random test is a good tracer of model quality for a broader range of magnitudes, and the faint extrapolation test is more sensitive to overfitting. During the model training, both testing methods should be taken into consideration. In the case of the classification, which achieves high accuracy, the faint extrapolation test can be given more importance. For redshifts, which are more difficult to fit at the faint data end, the extrapolation test might not show the full learning process, as illustrated by early minimums in QSO and galaxy redshift performance. When training the final inference models, we have to use the full magnitude ranges for training, so the extrapolation test is not available at that point, and we stop the model training based only on the results from the random test. Therefore, the best optimization approach during the experiments is to aim not only for the lowest error in a random test, but also for the lowest error in the extrapolation in the moment when the random error achieves its global minimum. This way, we can make sure that the final inference models, whose training is stopped based only on the random test, will also achieve good results at the faint data end.

Machine learning models can be modified in many ways which control the bias versus variance trade-off, in addition to the number of trees investigated in Fig.~\ref{fig:history_xgb}. In the case of ANNs, we tuned the number and size of layers, regularization, dropout, and learning rate. Some attempts at model optimization showed improvement in the results for both the tests, while the increased regularization usually led to better results only in the faint extrapolation case. For instance, once we reached the optimal network size for classification, using more layers or nodes per layer did not show any change in the random test, but led to deterioration in the faint extrapolation. Using only the randomly chosen subset may lead to a different set of parameters than when an extrapolation subset is also incorporated, and uncontrolled failure of estimation for the faint end. In case of incorrectly regularized models, such a failure can happen not only in extrapolation data, but also for the faintest magnitudes covered by the spectroscopic training data ($r \sim 22$ in our case). Thanks to both tests, we have the full picture of the bias versus variance trade-off and we can tune the models so that they perform well on both bright and faint data, and extrapolate to magnitudes fainter than available from spectroscopy. We consider this an important success of our approach.

We tested several ML strategies, and we conclude that two ANNs, one for classification and one for QSO redshifts, provide the best results overall\footnote{The final model parameters and ANN architecture can be found in the script \textit{models.py} in the github repository 
\url{https://github.com/snakoneczny/kids-quasars}.}. We find that a neural network model with multiple outputs for classification and redshifts, which would allow us to solve both problems at once, can be tuned to provide some improvement either for detection or redshift over two separate networks, but we did not manage to tune the network to simultaneously achieve the best results for both problems. It is due to both problems requiring different parameters. The specialized redshift models, trained either on galaxies or QSOs, are necessary for the best results, due to the differences between the two classes in the optimal model parameters, such as ANN size or regularization.

Table \ref{table:model_comparison} shows the results of the specialized redshift models. The redshift metrics (Section \ref{subsection_evaluation}) were calculated on two subsets of QSOs: true spectroscopic ones and our QSO candidates from photometric classification, as explained in Section \ref{subsection_evaluation}. In our previous work (N19), which dealt with classification only, we did not observe a significant difference between RF and XGB performance. In this work, we find distinct results between all the tested models, due to a more complex validation method and larger feature space, now extended by near-IR bands. In the random test, XGB performs best in classification, and ANN performs best in redshifts. The faint extrapolation test shows less agreement on which model is the best for classification, but the superiority of ANN for redshifts is more prominent. We find that XGBoost is the most robust and straightforward model for classification, while ANN is the best for a combined classification and redshift.

A mixed approach, where classification is performed with XGB and redshifts with ANN gives the best results on the random test, but worse results for QSO candidates in the faint test, due to different characteristics of both models resulting in fewer objects with both class and redshift assigned correctly. In the case of the faint extrapolation test and the subset of QSO candidates, the R\textsuperscript{2} deteriorates by 3 percentage points, while the standard deviation of $\delta z$ is higher by 0.03.

Artificial neural networks provide good extrapolation results for both classification and redshifts. The classification deteriorates by 3 percentage points in the faint extrapolation test, while the standard deviation of $\delta z$ is higher by 0.07 than in the random test.

\begin{figure*}
        \resizebox{\hsize}{!}{\includegraphics{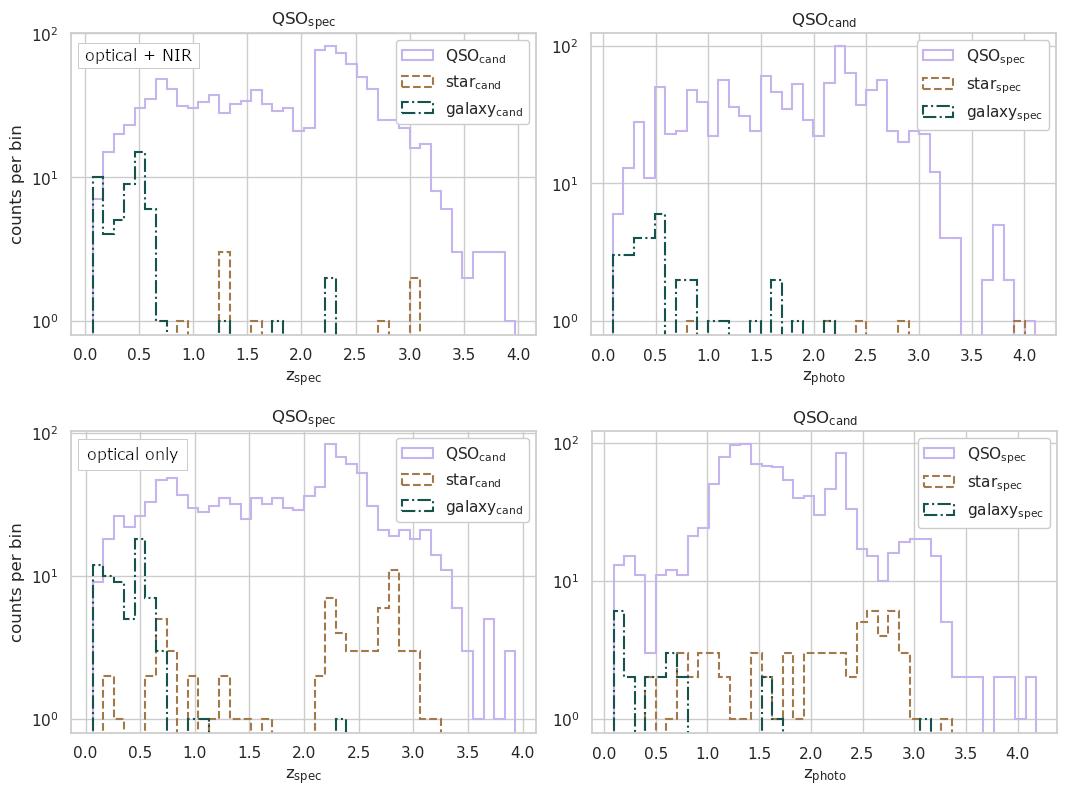}}
        \caption{QSO misclassification as a function of redshift. \textit{Top:} Using optical KiDS and near-IR VIKING features. \textit{Bottom:} Using only optical KiDS features. \textit{Left:} Spectroscopic QSOs and redshifts -- a test for completeness. \textit{Right:} QSO candidates and redshifts -- a test for purity.}
        \label{fig:qso_misclassification}
\end{figure*}

Quasar misclassification occurs mostly at low redshift (Fig. \ref{fig:qso_misclassification}), with AGNs which have extended hosts and are generally labeled as QSO by SDSS. This affects the completeness more than purity, as in broad-band optical and NIR photometry those AGNs are more similar to galaxies than to quasars. It is due to the spectra taken through fibers in the SDSS, and in case of galaxies with AGN, the fiber is centered on the nucleus. This allows resolved galaxies to be matched with a QSO template by SDSS, and be spectroscopically classified as quasars. The KiDS photometry, however, picks up the host galaxy light and does not allow one to see the emission lines, therefore such AGNs are classified as galaxies from imaging. We define quasars as all the objects labeled as QSO by the SDSS, and this  misclassification is a consequence of a mismatch in the QSO definitions between the spectroscopic and imaging surveys. Quasars at low redshifts with a low value of the stellarity index may additionally look more similar to extended galaxies for ML models. The quasar candidates consist of 96.9\% true quasars, 2.6\% galaxies, and 0.4\% stars. The bottom plots of Fig. \ref{fig:qso_misclassification} show results obtained using only the optical $ugri$ broad-bands. We observe misclassification with stars at the QSO redshift of $2 < z < 3$ (bottom left), and worse redshift estimates (bottom right), when only KiDS optical imaging is used, as studied previously in N19.

\begin{figure*}
        \resizebox{\hsize}{!}{\includegraphics{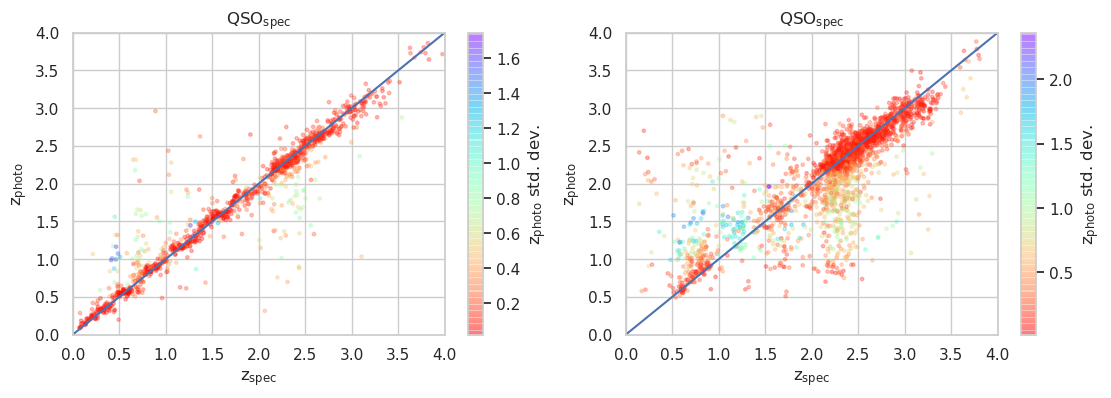}}
        \caption{Comparison of the spectroscopic and photometric redshifts for SDSS test-set quasars. \textit{Left:} Random test ($r < 21.3$). \textit{Right:} Faint extrapolation ($21.3 < r < 22$). The mean photo-z error for the random and faint test equals $0.009 \pm 0.12$ and $-0.0004 \pm 0.19$, respectively. Every redshift estimate is a Gaussian probability density function, the standard deviation of which represents the uncertainty (color coded).}
        \label{fig:qso_z}
\end{figure*}

Figure \ref{fig:qso_z} compares spectroscopic and photometric redshifts on the random and faint tests. The random test shows a well-fitted distribution and thus the modeled uncertainty increases for objects further from the diagonal. We observe some clustering of redshifts around several values in the random test, but we did not manage to establish whether it is due to the ML model or internal data characteristics. The outliers behave similarly also in spectroscopic measurements due to confusion between pairs of emission lines \citep[e.g.,][fig. 10]{Croom:2009}. The faint extrapolation test shows more scatter and more outliers. The aleatoric uncertainty, which we model with a Gaussian output layer, is related to the fact that objects which appear similar in photometry may have different redshifts. This model does not include the situation in which part of the feature space is not covered by data, and we would expect higher uncertainty for such estimations -- this case would relate to epistemic uncertainties. After several iterations of tuning the model with random and faint extrapolation tests, we managed to achieve useful uncertainties also for the faint extrapolation test, not covered by the training data.

As already mentioned, KiDS provides photometric redshifts for all cataloged galaxies, including quasars, and they are stored in the Z\_B column \citep{Kuijken:2019}. As these photo-z estimates were optimized for galaxies used for weak lensing studies, they are not expected to perform well for quasars in general. For comparison with our results, the mean error of the \textsc{bpz} estimates for the QSOs in the random test is $\delta z = -0.38 \pm 0.43$, while in the extrapolation\footnote{We note that as \textsc{bpz} is a template-fitting approach, its photo-z derivations are independent of the properties of training sets.}, $\delta z = -0.45 \pm 0.32$. The \textsc{bpz} redshifts for QSOs are significantly underestimated and much less precise than our estimates: Their scatter is 3.5 times higher in the random test, and 1.7 times higher in the faint extrapolation test, in comparison to our results.

\begin{table*}
        \caption{ANN results on MASK flagged objects in the random ($r < 21.3$) and faint extrapolation ($21.3<r<22$) tests. Brackets show differences to corresponding ANN results from Table \ref{table:model_comparison}.}
        \centering
        \begin{tabular}{c c c c c}
                \hline\hline
                test & purity & recall & $\delta z$ for true QSOs & $\delta z$ for QSO candidates \\
                \hline
                random & 96.41\% (-0.49\%) & 93.98\% (-0.74\%) & 0.004 (-0.005) $\pm$ 0.13 (+0.01) & 0.019 (+0.000) $\pm$ 0.25 (+0.06) \\
                faint extrap. & 94.36\% (-2.18\%) & 88.28\% (-2.55\%) & 0.01 (+0.01) $\pm$ 0.22 (+0.03) & 0.07 (+0.03) $\pm$ 0.41 (+0.04) \\
                \hline
        \end{tabular}
        \label{table:mask_results}
\end{table*}

The KiDS DR4 catalog provides a MASK flag indicating possible flux contamination from issues such as star halo, globular clusters, ISS, etc. We observe stability of the estimations in a random test on objects with such contamination. To verify this, we evaluated ANNs on the objects flagged with any MASK bit (Table \ref{table:mask_results}). The results are stable in the random test and show some deterioration in the extrapolation test. We always include all masked objects in the training, so the models can learn how to process them, and the associated additional noise helps in regularization.

\begin{figure*}
        \resizebox{\hsize}{!}{\includegraphics{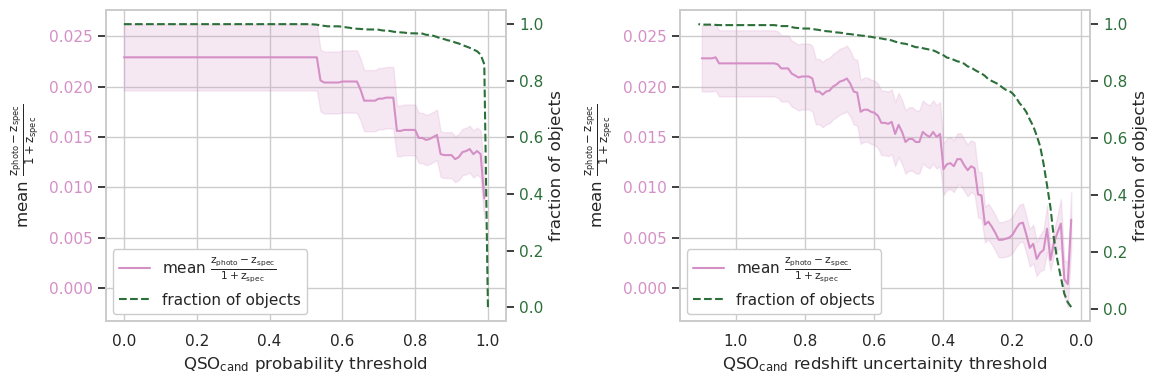}}
        \caption{QSO photometric redshift errors as a function of thresholds in QSO probability (\textit{left panel}) and model photo-z uncertainty (\textit{right panel}). An increasing minimum classification probability yields better redshift estimations at a small cost in completeness. Low uncertainty estimations further increase redshift reliability at a cost of removing more objects.}
        \label{fig:cleaning}
\end{figure*}

Classification and redshift results can be improved by limiting the sample to objects with higher classification probabilities or lower redshift uncertainties (Fig. \ref{fig:cleaning}). We consider the classification probability limits as the primary way to calibrate the catalog's purity-completeness trade-off, while the uncertainties can be used to achieve the necessary redshift precision.

\subsection{Final catalog properties}
\label{subsection_final_catalog}

We applied the trained ML models to 45M objects of the KiDS DR4 inference data, and we find a total of 3M QSO candidates, excluding the unsafe inference subset. In the final model training, we used the whole range of magnitudes of the training set, as well as a randomly selected validation sample. We employed the same set of values of hyper-parameters as determined in the experiments which included the faint extrapolation test, and we only picked a new number of epochs based on new learning histories with a randomly selected test sample.

\begin{figure}
        \resizebox{\hsize}{!}{\includegraphics{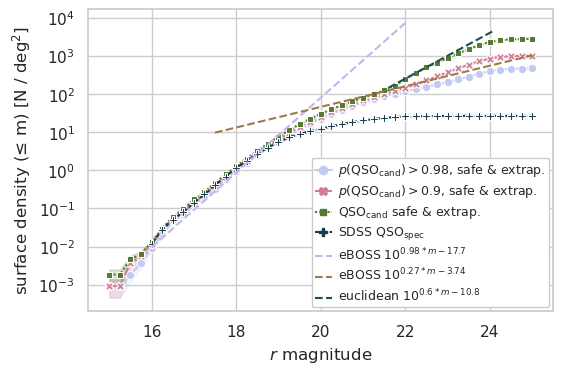}}
        \caption{QSO number counts of SDSS spectroscopic QSOs and KiDS QSO candidates ($\rm QSO_{cand}$) at progressing classification probability cuts, excluding the unsafe inference subset. The dashed lines show eBOSS predictions fitted with a broken power law. The SDSS spectroscopic QSOs are complete to $r < 19$. KiDS QSO candidates without a probability cut are too numerous at $r > 21.5$ due to misclassification, and they follow standard Euclidean number counts. A cut at $p(\rm QSO_{cand}) > 0.9$ gives a complete catalog in the safe subset ($r < 22$). A cut at $p(\rm QSO_{cand}) > 0.98$ provides expected number counts up to $r \lesssim 24$.}
        \label{fig:number_counts}
\end{figure}

In Figure \ref{fig:number_counts} we compare the number counts of QSO candidates ($\rm QSO_{cand}$) in the safe and extrapolation subsets to the predictions from the eBOSS survey \citep[table 7 from][]{Palanque:2016}. We fit the eBOSS predictions with a broken power law. Our analysis suggests that two cuts on the photometric QSO probability match the expected numbers: $p(\rm QSO_{cand} > 0.9$ for the safe magnitude range ($r < 22$) and $p(\rm QSO_{cand}) > 0.98$ for the extrapolation. The fit of the QSO number counts to eBOSS predictions is reliable for $r < 23.5$, where the extrapolation subset is complete (Fig. \ref{fig:inference_subsets}). We do not observe the expected decrease in the QSO number counts at $r > 23.5$, which should result from reaching the completeness limit of the extrapolation subset. This suggests increased impurity of the QSO candidates in that range. The possible unreliability of the classification at $r > 23.5$ was already suggested by the t-SNE visualization in Fig. \ref{fig:t-SNE_subsets}.

\begin{figure}
        \resizebox{\hsize}{!}{\includegraphics{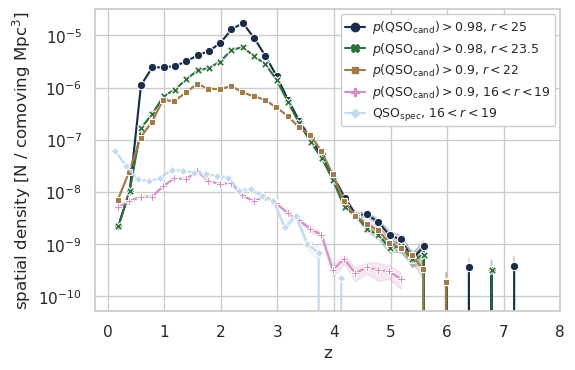}}
        \caption{Spatial number densities, excluding the unsafe inference subset, for KiDS QSO candidates. Two bottom lines compare the KiDS QSO candidates to the SDSS spectroscopic QSOs at the SDSS completeness range of $16 < r < 19$. The three upper lines show the final QSO catalog at progressing magnitude limits with the suggested probability cuts. We chose a magnitude limit for the middle line at $r < 23.5$, as above this limit the distribution of QSO candidates gains another peak at redshift $z < 1.5$.}
        \label{fig:spatial_density}
\end{figure}

Figure \ref{fig:spatial_density} shows spatial number densities for KiDS QSO candidates based on the photometric redshifts and for SDSS spectroscopic QSOs based on the spectroscopic redshifts. We accounted for the $V_{\rm max}$ correction, taking the KiDS magnitude limit $r=25$ and assuming the WMAP9 \citep{Hinshaw:2013} cosmology. The distribution is expected to peak at $z\sim2$ - 3 and then follow an exponential decrease \citep{Fan:2006}. Based on the SDSS spectroscopic QSO number counts (Fig. \ref{fig:number_counts}), we estimated its completeness to be $r < 19$. We observe some differences between KiDS photometric and SDSS spectroscopic QSO densities at this limit. The QSOs missing at low redshifts are due to the previously discussed misclassification with galaxies (Fig.~\ref{fig:qso_misclassification}). At the faintest end ($r>23.5$), on the other hand, the photo-z-based density displays an additional peak at $z < 1$ for the suggested $p(\rm QSO_{cand}) > 0.98$. This is due to apparently faint galaxies classified by our model as QSOs and assigned redshifts lower than one. This conclusion agrees with the number counts indicating a QSO impurity at $r > 23.5$.

\begin{table*}
        \caption{Number of photometrically selected QSOs in our catalog at progressing magnitudes with the suggested probability cuts (bold), excluding the unsafe inference subset. At fainter magnitudes, a higher probability threshold is required for robustness. The cuts give smaller subsets of QSO candidates and increase the purity.}
        \centering
        \begin{tabular}{r c c c}
                \hline\hline
                & safe $r < 22$ & safe \& extrap. $r < 23.5$  & safe \& extrap. $r < 25$ \\
                \hline
                $\rm QSO_{cand}$ & 266k (100\%) & 1.6M (100\%) & 3M (100\%) \\
                $p(\rm QSO_{cand}) > 0.90$ & \textbf{158k (59\%)} & 637k (39\%) & 1.1M (36\%) \\
                $p(\rm QSO_{cand}) > 0.98$ & 127k (48\%) & \textbf{311k (19\%)} & 507k (17\%) \\
                \hline
        \end{tabular}
        \label{table:catalog}
\end{table*}

Table \ref{table:catalog} summarizes the number of QSOs in the final catalog at progressing magnitude limits -- thus reliability limits -- and the suggested probability cuts. According to the number counts and spatial number densities, the QSO classification and redshift estimations should be reliable up to $r < 23.5$. At $r > 23.5$, the classification provides excessive number counts, and the photometric redshifts suggest misclassification with galaxies. The forthcoming DESI and planned 4MOST QSO surveys could help verify these finding, as they will include QSOs fainter than SDSS and will overlap with KiDS.

\begin{figure*}
        \centering
        \resizebox{\hsize}{!}{\includegraphics{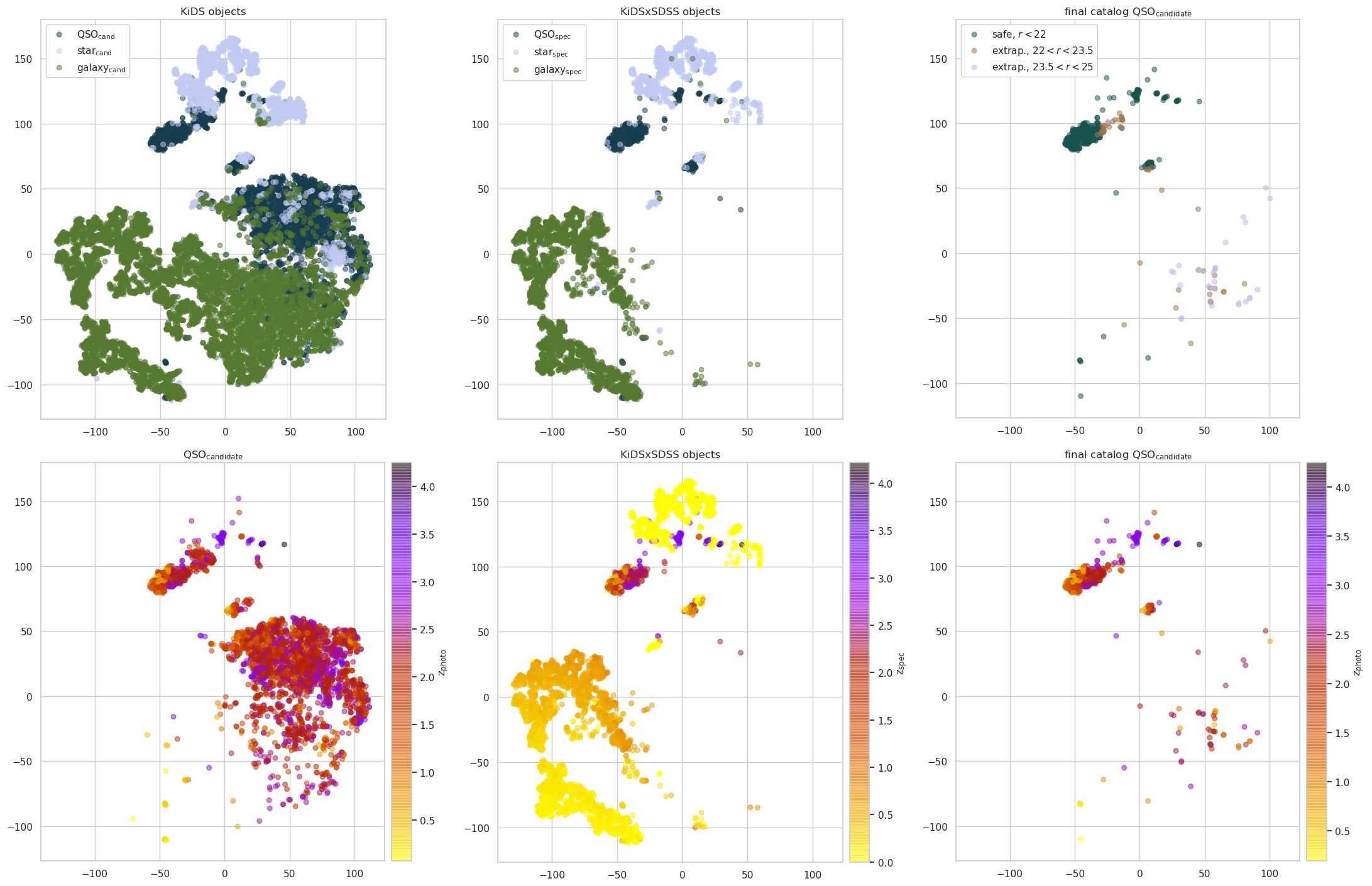}}
        \caption{t-SNE projections. Top: Classification. Bottom: Redshifts. Left: Raw output from the ML models for all the inference subsets. Center: Spectroscopic SDSS distributions. Right: Final QSO catalog at progressing magnitudes with the corresponding probability cuts, excluding the unsafe inference. The visualizations were made on a subset of 12k objects, thus actual object density at any part of the feature space is 3.8k times higher.}
        \label{fig:t-SNE_results}
\end{figure*}

We visualized the outputs from the ML models, compared it to the spectroscopic information, and show the final catalog properties for the inference subsets and suggested probability cuts using t-SNE in Fig. \ref{fig:t-SNE_results}. The main spectroscopic QSO group is accurately covered with photometric classification and redshifts. In the close extrapolation, the predictions appear as a regular extension of the main QSO group, which qualitatively confirms the success of our approach. The decision to separate out the unsafe inference subset is confirmed, as we observe the distributions of all three classes overlapping in the corresponding part of the feature space. The estimations for fainter magnitudes could be used to look for QSOs at the highest redshifts or to select candidates for follow-up spectroscopy.

\subsection{Gaia parallaxes}
\label{subsection_gaia}

\begin{figure}
        \resizebox{\hsize}{!}{\includegraphics{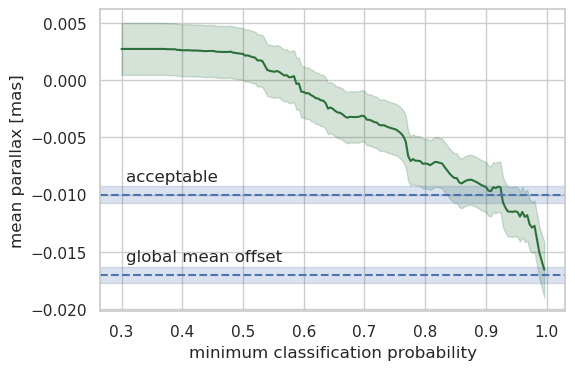}}
        \caption{Mean parallax for KiDS DR4 QSO candidates as a function of minimum classification probability. The Gaia observations have a global mean offset, which is imprinted in the QSO mean parallax distribution. The offset for SDSS spectroscopic QSOs equals $-0.017 \pm 0.001$ mas (standard error on the mean). We calculated the acceptable offset based on star and galaxy contamination estimated in the experiments. It equals $-0.01 \pm 0.0015$ mas.}
        \label{fig:gaia_mean_proba}
\end{figure}

We cross-matched the QSO candidates identified here with Gaia DR2 \citep{Gaia:2018} to estimate the star contamination. A clean set of QSOs is expected to have a global mean parallax offset of $-0.029$ mas \citep{Lindegren:2018}. This value was calculated by removing incorrectly measured high parallaxes for SDSS QSOs. Following the same procedure for KiDS, QSO candidates would remove the star contamination, which we want to measure. Instead, we calculated a less precise mean offset for SDSS QSOs in a high precision sample with parallax and proper motion errors smaller than 1 mas. This offset equals  $-0.017$ mas, which is smaller in absolute terms than the official Gaia measurement.

The QSO candidates in the safe inference subset show a mean parallax offset of 0.003 mas, and this goes down at the progressing minimum classification probability (Fig. \ref{fig:gaia_mean_proba}). This assessment is based on a cross-match between our catalog and the Gaia high precision sample mentioned above, which yields 1.63M objects: 1.61M (98.7\%) classified photometrically as stars, 20k (1.2\%) as QSOs, and 1k (0.1\%) as galaxies. The test is limited to the Gaia magnitude $G<21$, which corresponds to $r \lessapprox 20$. We then calculated an \quotes{acceptable offset} from a sample of the three spectroscopic classes, with the size of each class corresponding to the contamination of QSO candidates with stars and galaxies derived from the experiments: 96.9\% QSOs, 2.6\% galaxies, and 0.4\% stars (Fig. \ref{fig:qso_misclassification}). The minimum QSO photometric probability suggested by this test is $p(\rm QSO_{cand}) = 0.9$. This cut, which was obtained from the more precise test at $r \lessapprox 20$, agrees with the cut for the safe inference subset at $r < 22$ derived from the number counts.

\subsection{Comparison with other QSO catalogs}
\label{subsection_external_test}

\begin{figure}
        \resizebox{\hsize}{!}{\includegraphics{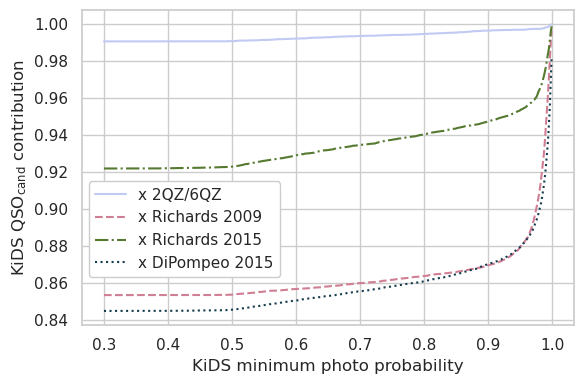}}
        \caption{Proportion of KiDS DR4 QSO candidates in cross-matches with other QSO catalogs as a function of KiDS minimum photometric classification probability.}
        \label{fig:ext_qso_proba}
\end{figure}

We find good agreement with other QSO catalogs overlapping with the KiDS DR4 footprint (Fig. \ref{fig:ext_qso_proba}). Additional ground-truth samples, which were not used in the training, provide a good test of ML estimations. We used additional QSO catalogs built from different datasets and with different methodologies than ours. Those involve one spectroscopic catalog, 2QZ / 6QZ \citep[][hereafter 2QZ]{Croom:2004}, and the following three photometric ones by \cite{Richards:2009a, Richards:2015, DiPompeo:2015}, hereafter R09, R15, and DP15, respectively. 2QZ includes QSOs, stars, and galaxies confirmed with spectroscopy, while the photometric catalogs are probabilistic, based on a selection from SDSS (R09) and SDSS+WISE (R15 \& DP15). DP15 publishes the whole range of QSO probabilities, which we limited to higher than 70\%, according to the distribution with shows a minimum number of objects at this value. 2QZ, being spectroscopic, can be used as ground truth and confirms high QSO purity and completeness of our sample: 98.2\% three class accuracy, 98.6\% QSO purity, and 99.4\% QSO completeness. We note, however, that as 2QZ sources are on average brighter than those from the SDSS QSO catalog, these numbers should not be taken as measurements of the overall performance of our classification.

\section{Discussion}
\label{section_discussion}

\subsection{Main findings}

In this paper we employed supervised ML models to identify QSOs in KiDS DR4 and evaluate their redshifts. We found 158k QSO candidates with a minimum classification probability of $p(\rm QSO_{cand}) > 0.9$ at $r < 22$, and a total of 311k QSO candidates with $p(\rm QSO_{cand}) > 0.98$ for $r < 23.5$, that is to say in the extension to the close extrapolation data. The far extrapolation at $r < 25$ provides a total of 507k QSO candidates at $p(\rm QSO_{cand}) > 0.98$. The catalog of QSOs is well designed for extrapolation, with the reliability regions derived from visualizations, and probability thresholds calibrated via a series of tests. Based on the SDSS QSO test sample, the purity of the catalog is 96.9\%, and completeness is 94.7\% for $r < 22$. The extrapolation by $\sim$0.7 magnitude lowers the purity by 0.4 percentage points and the completeness by 3.9 percentage points. The average redshift error in terms of $(z_{\rm photo} - z_{\rm spec}) / (1 + z_{\rm spec})$ equals $0.009 \pm 0.12$ for $r<22$, with its scatter increasing to $-0.0001 \pm 0.19$ in the extrapolation ($r<23.5$).

We found that the traditionally adopted testing method, based on randomly selected samples of objects, was insufficient to tune the bias versus variance trade-off. A faint-end test is necessary for the proper extrapolation of both classification and redshifts, but also important for appropriate tuning and inference on the bright end data. This approach towards ML model calibration and the satisfactory extrapolation results are the main novelty aspects of our work. Thanks to the faint extrapolation test, we also obtain useful redshift uncertainties in the extrapolation data, even though we used the Gaussian output layer to model aleatoric uncertainty. Otherwise, we would expect the aleatoric uncertainty to fail in the part of the feature space not covered by the training data.

The addition of the near-IR VIKING bands, which were not available in the KiDS DR3 on which N19 was based, provided crucial information for QSO redshifts and helped us to distinguish stars from QSOs at redshifts of $2 < z < 3$. The most important bands for QSO redshifts, according to our experiments, are the near-IR $ZK_s$, which are the two extreme bands covered by VIKING. This suggests that it is the span of the infrared wavelengths that is relevant here. We found it important to use both magnitude differences (colors) and magnitude ratios. Interestingly, colors and ratios constructed from the same magnitude pairs had a different importance for the ML models. What is more, the ratios were in fact more common than colors among the most important features used by XGBoost for classification and QSO redshifts. This experimental analysis could be further perfected with proper fine tuning of the models trained using a no-ratio feature set in order to draw the final conclusions. Additionally, possible further experiments may involve more custom feature engineering based on flux values in order to find the most robust photometric features.

The comparison of ML models also shows clear trends: XGB performs better at classification, while ANN provides a better redshift estimation, that is to say it works better for regression. Many astronomical papers report no such differences, which was also the case in our previous work (N19). We uncovered these differences as more features are available from the VIKING imaging, which allowed us to obtain better results with more sophisticated classification models such as XGB. The superiority of ANN for regression is largely due to its better performance in extrapolation, not only in feature space, but also in higher values of the estimated photo-zs. The models tuned for both random and faint extrapolation tests are also less overfitted and show real differences between their characteristics.

We successfully supported our analysis with t-SNE projections of high-dimensional space onto 2D, instead of the standard color-color plots. The visualizations helped us to derive a reliable inference subset at close extrapolation, which was possible by verifying the location of these extrapolation data with respect to the feature space known from spectroscopic classification. We also used the projections to test different feature sets. The distribution of spectroscopic classes on the t-SNE plots allowed us to initially assess the reliability of feature engineering, without even training a supervised model. Last but not least, the visualizations helped us understand where the classification fails due to overlapping distributions between various object classes in the feature space.

\subsection{Relation to other work}

Most of the QSO classification and redshift estimation studies are not directly comparable due to the results depending on available bands, survey brightness, size of the training sample, and different definitions or detection schemes of AGNs and QSOs in spectroscopy and photometry. To ensure both high purity and completeness using color-color cuts, one has to model the data with many distributions or build a set of decision boundaries \citep[e.g.,][]{Richards:2002}. On the other hand, ML allows us to build the most complicated decision boundaries in an automatic way,  while simultaneously optimizing both purity and completeness. The power of ML approaches comes with the danger of possible overfitting. This problem is usually not addressed in the ML analyses, and the results on faint end data, which are most affected by overfitting, are rarely reported \citep[e.g.,][]{Hausen:2020}. As far as we know, our results for data fainter by one magnitude than the reach of the training data -- completeness lower by 3\% and redshift scatter increased by 0.07 in comparison to the regime covered by the training -- are reported for the first time. This outcome challenges the way that ML models are usually optimized and applied on the faint data end. For other problems, other data characteristics can be used to obtain extrapolation tests, for example, the high and low mass end for galaxy cluster mass estimation, the number of objects in n-body problems, cosmological parameters not available during training in cosmological problems, etc.

Our work is the first in which the simultaneous selection of QSOs from photometry and evaluation of their photometric redshifts is performed for samples selected from the KiDS+VIKING catalog. In a recent study, \citet[][L20]{Logan:2020} performed a classification and redshift estimation in KiDS DR4, but on a smaller subset of 2.7M objects selected over 200 deg$^2$ with the additional requirement of available detections in the WISE mid-IR bands. That classification was done with unsupervised hierarchical density-based spatial clustering of applications with noise \citep[HDBSCAN,][]{McInnes:2017}, redshift estimation with RF, and feature engineering with principal component analysis \citep[PCA,][]{Pearson:1901}. A quantitative comparison of our catalogs with respect to experimental results on SDSS data is not possible due to different train and validation strategies. We have, however, performed a qualitative comparison using the full training data from the L20 catalog. The classification results are different as L20 uses an unsupervised algorithm, which does not allow for a  completeness that is as high as our supervised approach. We find our photo-zs to be more precise on average, but L20 photo-zs are more robust at the faint end.

As already mentioned in the Introduction, we addressed the QSO selection problem in KiDS in our previous work, N19, where we applied an ML classification to the DR3 $ugri$ photometry. In that study, we employed the RF algorithm and reported 91\% purity and 87\% completeness for QSOs. In the present work, most of the improvement in classification comes from adding the NIR bands, which allowed us to correctly classify QSO at $2.5 < z < 3$, where they are similar to stars in the \textit{ugri} broad-bands. Additionally, two significant improvements were made: We now provide QSO photometric redshifts, and publish estimations for objects fainter than the training data, with models tuned for extrapolation.

Another related work is the KiDS Strongly lensed QUAsar Detection project \citep[KiDS-SQuaD;][]{Spiniello:2018, Khramtsov:2019}, aimed at finding strongly gravitationally lensed quasars in the KiDS data. This latter paper in particular describes the KiDS Bright EXtraGalactic Objects catalog (KiDS-BEXGO), constructed from DR4 and including about 200k sources identified as QSOs based on an application of the CatBoost gradient boosting ensemble algorithm \citep{Prokhorenkova:2018}. The BEXGO catalog is optimized for the lowest possible star contamination at a cost of reduced completeness, and it is limited to $r < 22$. The results of an ML QSO identification are not directly comparable between our work and that of \cite{Khramtsov:2019}, as in the latter the QSOs are defined as point-like objects, and any AGNs with a visible galaxy host had been removed from the training data, unlike in our case. We have kept QSOs, which appear extended in our training data, as such sources provide useful information on the relation between QSOs and galaxies at low redshifts. It might have a vital outcome on the final predictions and possibly makes both catalogs different.

Furthermore, the dataset constructed by \cite{Khramtsov:2019} is aimed to carry out the specific purpose of QSO strong lensing, which requires the highest possible purity of the catalog. The approach that we have taken, on the other hand, is to obtain the most optimal purity-completeness trade-off, which requires ML models to be properly tuned to the given problem and data. A required level of purity or completeness can then be acquired a posteriori by properly calibrating the catalog, in particular by applying appropriate cuts on the probability that a given source is a QSO.

We envisage that our catalog of QSOs can have versatile applications in studies related to AGNs or LSS, as it is optimized solely for QSO identification without outside requirements. The availability of robust photometric redshifts with uncertainty estimates for the QSOs contained in our catalog is expected to prove especially useful in approaches where \quotes{tomographic} dissection of the LSS is done, such as cross-correlations with various backgrounds.

In this work, we trained the ML models to perform a full three-class classification on both extended and point-like objects. If instead one was not interested in AGNs with resolved galaxy hosts, but only point-like QSOs at higher redshifts, then based on the finding of our work, we suggest to train the ML classifier only on point-like objects -- for example, those with the stellarity index higher than 0.8 -- and apply only QSO versus star classification. Such a model is easier to train and interpret, and visualizations of the relevant data are simpler to understand than in the full three-class problem including both extended and point sources.

\subsection{Limitations and possible improvements}

We consider our approach towards the inference at the faint data end, which involves tuning the model based on a faint extrapolation test, as the most optimal as far as the current supervised ML models are concerned. However, a reliable test of our predictions outside of the magnitude coverage of spectroscopic samples is not possible, and at present KiDS does not overlap with any wide-angle samples providing sufficient numbers of spectroscopic QSOs beyond $r > 22$. This situation will likely improve in the coming years thanks to the already ongoing DESI \citep{DESI:2016} and planned 4MOST \citep{Merloni:2019,Richard:2019} QSO surveys, which will largely overlap with KiDS.

The random and faint extrapolation tests require an interpretation, which depends on the problem complexity and robustness of the inference at the faint end. When determining the appropriate value of a given model parameter, for example the number of epochs or trees, one might obtain ambiguous results, such as a range of acceptable values rather than one best value. This adds to the complexity of model optimization. The results on faint end extrapolation are reported to have a high impact on the estimation reliability \citep[e.g.,][]{Shu:2019, Clarke:2020, Logan:2020}. We achieved satisfactory extrapolation results in $r < 23.5$, which is 1.5 magnitude larger than the SDSS limit. Our results are robust, because we not only find a limit at which the results diverge from expectations, but also make sure that the results are adequate for data brighter than this limit, $r < 23.5$ in our case.

The biggest source of incompleteness in our catalog comes from removing objects with at least one band missing out of the nine available. This decreases the size of the KiDS inference data by 55\%, from 100 million to 45 million. The requirement of $u$-band detections may lower the completeness of QSOs at $z\gtrsim2$. When looking for such high-$z$ QSOs, one would have to perform a classification and redshift estimation using only the redder bands. The possible addition of red QSOs to the training may result in a higher QSO density at high redshifts, at the cost of limiting the feature space.

Another source of incompleteness is the removal of 13 million of the faintest objects for which the SExtractor morphological classifier CLASS\_STAR fails. At $r > 23.5$, the unsafe subset constitutes a large fraction of all KiDS objects (65\%) and dominates at $r > 24$ (81\%) (Fig. \ref{fig:inference_subsets}). As the stellarity index is in fact one of the most important features for the classification (Fig. \ref{fig:features_xgb}), its inaccuracy  at the faint data end may account for the limit of reliable extrapolation, which is $r < 23.5$.

We plan several steps in order to further increase the catalog's completeness and interpretability. The missing data problem can be solved with either straightforward methods, such as assigning some specific values to the missing features, for example, zeros or mean values, or more sophisticated approaches such as predicting the missing values or using models designed to work with missing features \citep[e.g.,][]{Smieja:2018}. It might also prove necessary to skip the shape classifiers for the faint end estimations. The redshift uncertainties require epistemic uncertainty modeling in order to be fully useful in the extrapolation range of  $r > 22$. This can be implemented in ANN with, for example,  variational layers of Tensorflow, which represent each weight as a probability distribution.

It is possible to validate the faint-end predictions by fitting an SED to the QSO candidates in the catalog, using the estimated photo-zs as input to SED fitting. This will allow us to physically interpret the predictions and find the physical reasons for some of the model failures. Furthermore, this could be the best way of validating the estimations at the faint magnitude end by evaluating how physically acceptable the QSO SED fits are.

Dedicated spectroscopic observations might be yet another way of validating the estimations at extrapolation. They would allow us to determine more precisely the limit of reliability of our predictions at $r \approx 23.5$. It would be interesting to also probe the faintest objects to understand how the estimations cover the unsafe inference subset and find what is the actual portion of real QSOs in our selection in the faintest end. If the results are positive enough, this would show that the ML models optimized for the extrapolation can also serve as a method of candidate selection for follow-up spectroscopy in such faint data.

In this work we have shown how artificial intelligence can be successfully used to process large amounts of astronomical data. The wide-angle KiDS DR4 catalog of 253k QSO candidates with reliable photometric redshifts can be used in both AGN and LSS studies, and our work addresses important aspects for any other application of ML in astronomy. As we have demonstrated, well-designed inference models can be pushed to the limits and give reliable results even beyond the coverage of the training sets. The interested readers can test the approach of validation on the faint data proposed in this work in their own inference schemes, and compare what differences it brings to parameter optimization. This work, and ML processing in general, is important in a view of the upcoming large surveys such as the Rubin Observatory LSST or Euclid. Those new endeavors will provide unprecedented vast amounts of data much fainter than the current spectroscopic surveys, and also going deeper than most of the current wide-angle imaging datasets, which will require robust big data processing. Carefully designed, intepretable, and well-tested ML models can provide reliable and trustworthy results. We believe that the framework developed here is one step towards meeting the demands of these future missions.

\begin{acknowledgements}

We would like to express our gratitude to Sotiria Fotopoulou and Natasha Maddox for providing useful comments on the paper.

This research was supported by the Polish Ministry of Science and Higher Education through grant DIR/WK/2018/12.
SJN is supported by the Polish National Science Center through grant UMO-2018/31/N/ST9/03975.
MB is supported by the Polish National Science Center through grants UMO-2018/30/E/ST9/00698 and UMO-2018/31/G/ST9/03388.
AP is supported by the Polish National Science Center through grant UMO-2018/30/M/ST9/00757.
MA acknowledges support from the European Research Council under grant number 647112.
AD acknowledges ERC Consolidator Grant (No. 770935).
BG acknowledges support from the European Research Council under grant number 647112 and from the Royal Society through an Enhancement Award (RGF/EA/181006).
CH acknowledges support from the European Research Council under grant number 647112, and support from the Max Planck Society and the Alexander von Humboldt Foundation in the framework of the Max Planck-Humboldt Research Award endowed by the Federal Ministry of Education and Research.
HH is supported by a Heisenberg grant of the Deutsche Forschungsgemeinschaft (Hi 1495/5-1) as well as an ERC Consolidator Grant (No. 770935).
KK acknowledges support from the Royal Society and Imperial College.

\textit{Author Contributions:} All authors contributed to the development and writing of this paper. The authorship list is given in two groups: the lead authors (SJN, MB, AP), followed by an alphabetical group of those who have either made a significant contribution to the data products, or to the scientific analysis.

\end{acknowledgements}

\bibliographystyle{aa}
\bibliography{mybib}

\begin{appendix}
\section{Data products}

\begin{table}
        \caption{Columns provided in data products.}
        \centering
        \begin{tabular}{l l}
                \hline\hline
                Label & Description \\
                \hline
                ID & ESO ID \\
                RAJ2000 & Centroid sky position right ascension (J2000) \\
                DECJ2000 & Centroid sky position declination (J2000) \\
        MAG\_GAAP\_r & \textit{r}-band GAaP magnitude with optimal MIN\_APER (extinction corrected) \\
        CLASS\_STAR & SExtractor star-galaxy classifier\\
        MASK & 9-band mask information \\
        \{CLASS\}\_PHOTO & Probability that the source is in one of the three classes: GALAXY, QSO, STAR \\
        CLASS\_PHOTO & Object class with the highest probability \\
        Z\_PHOTO\_QSO & Photometric redshift for quasars \\
        Z\_PHOTO\_STDDEV\_QSO & Uncertainty of photometric redshift for quasars \\
        SUBSET & ML inference subset (Section \ref{subsection_inference_subsets}). Values: safe, extrapolation, unsafe. \\
        \hline
        \end{tabular}
        \label{table:catalog_columns}
\end{table}

Data are available at: \url{http://kids.strw.leidenuniv.nl/DR4/quasarcatalog.php}. Table \ref{table:catalog_columns} describes the data columns. Here we provide only a subset of the KiDS columns, the rest can be obtained by cross-matching with the full KiDS DR4 data by ID.

\subsection{Catalog of QSO candidates}

Filename: KiDS\_DR4\_QSO\_candidates.fits \\
File size: 110 MB \\
Number of objects: 1,095,711 \\
Data limited to:
\begin{itemize}
    \item 9-band detections
    \item $r < 25$
    \item CLASS\_STAR < 0.2 or CLASS\_STAR > 0.8
    \item $p(\rm QSO_{cand})$ > 0.9
\end{itemize}
Possible values for the inference subset: safe, extrapolation. Suggested cut for the extrapolation subset: $r < 23.5$ and $p(\rm QSO_{cand}) > 0.98$ (Table \ref{table:catalog}).

\subsection{Catalog of all machine learning estimates}

Filename: KiDS\_DR4\_all\_ML\_estimates.fits \\
File size: 5.5GB \\
Number of objects: 45,469,955 \\
Data limited to 9-band detections. \\
Possible values for the inference subset: safe, extrapolation, and unsafe.

\end{appendix}

\end{document}